\documentclass[preprint,aps,amsmath,amssymb,nofootinbib,12pt]{revtex4}
\usepackage{epsfig}
\usepackage{graphicx,subfigure}
\usepackage{slashed}
\usepackage{float}

\usepackage{multirow,dcolumn,color}
\usepackage{amsmath}

\usepackage{amssymb}
\usepackage{enumerate}
\usepackage{verbatim}
\usepackage[sc]{titlesec}

\graphicspath{{fig/}}

\def\beq{\begin{equation}}
\def\eeq{\end{equation}}
\def\bea{\begin{eqnarray}}
\def\eea{\end{eqnarray}}

\begin{document}
\title{Searching for heavy vector-like B quark via pair production in fully hadronic channels at the CLIC}
\author{Shuo Yang}
\thanks{shuoyang@lnnu.edu.cn}
\affiliation{Department of Physics, Liaoning Normal University, Dalian 116029, China }
\affiliation{Center for theoretical and experimental high energy physics, Liaoning Normal University,Dalian 116029, China}
\author{Yi-Hang Wang}
\affiliation{Department of Physics, Liaoning Normal University, Dalian 116029, China }
\author{Peng-Bo Zhao}
\affiliation{Department of Physics, Liaoning Normal University, Dalian 116029, China }
\author{Ji-Long Ma}
\affiliation{Department of Physics, Liaoning Normal University, Dalian 116029, China }

\begin{abstract}
Vector-like quarks (VLQs) are introduced in many new physics senarios beyond the Standard Model (SM) to address some problems faced by SM. In this paper, we  explore the pair production of TeV-scale vector-like B quark (VLQ-$B$) at the future 3 TeV Compact Linear Collider (CLIC)  in simplified effective lagrangian framework. We consider the decay modes of $B\rightarrow bZ$ and $B\rightarrow bh$ followed by hadronic decay of $Z$ and $h$ bosons.  The large mass of  VLQ-$B$ will induce highly boosted  bosons $Z$ or $h$ which are more likely to form as fat-jets.  By performing a rapid detector simulation of the signal and background events and clustering the jets with a large radius R, signal-background analyses are carried out. And the exclusion limit at the 95\% confidence level and the 5$\sigma$ discovery prospects are obtained with an integrated luminosity of  5$\text{ab}^{-1}$.
\end{abstract}

\maketitle

\section{Introduction}
Extra quarks are supposed to offer solutions to some fundamental issues of the standard model on electroweak symmetry breaking and mass generation of
fundamental particles in many new physics models~\cite{DeSimone:2012fs,ArkaniHamed:2002qy,Han:2003wu,Chang:2003vs,Agashe:2004rs,He:1999vp,Wang:2013jwa,He:2001fz,He:2014ora}.
An extra fourth generation of SM-like quarks suffer serious electroweak precision constraints and Higgs search constrains~\cite{He:2001tp,Hung:2007ak,Hashimoto:2010at,Buras:2010pi,Denner:2011vt,Djouadi:2012ae}.
Unlike the quarks of the Standard Model and the sequential fourth generation quarks,  the left-handed and right-handed chiral components of the vector-like quarks (VLQs) transform under the same representation of the SM gauge group. This means that VLQs  don't  get their mass from interactions with the Higgs field,  and hence are not excluded by LHC measurements involving the Higgs boson. According to the charge assignments under the SM electroweak gauge group, VLQs can be electroweak singlets [T,B], doublets [ (X,T),~(T,B) or (B,Y) ] or triplets [ [X,T,B] or [T,B,Y] ].  Considering single or pair production of VLQs,  ATLAS and CMS has carried out extensive searches and analyses for these VLQs (For recent reviews, see Refs. ~\cite{ATLAS:2024fdw,CMS:2024bni}).  No evidence for these particles has been found and this has led to lower bounds on their masses above the TeV~\cite{ATLAS:2024fdw,CMS:2024bni,Benbrik:2024fku,ATLAS:2022hnn}.

But this is not the end of the story of  VLQs. Current experimental searches for VLQs have mainly considered their decays into Standard Model particles. For example, the ATLAS Collaboration presented a search for the pair production of VLQs in which at least one of the VLQs decays into a leptonically decaying $Z$ boson and a third-generation quark using the full Run 2 dataset corresponding to 139 $\text{fb}^{-1}$ ~\cite{ATLAS:2022hnn}.  Considering only three possible decay modes of $B\rightarrow tW, bh, bZ $,  the lower limits on the masses of  VLQ-$B$ (with charge -$1/3$)  are 1.20~TeV and 1.32 TeV for the weak-isospin singlet model and doublet model, respectively~\cite{ATLAS:2022hnn} .

However,  VLQs can also decay to new channels in many variant new physics scenarios predicting additional scalars. Recently, exotic decays of the VLQs in different set-ups with rich collider signatures have been considered in the literature~\cite{Banerjee:2024zvg,Aguilar-Saavedra:2017giu,Das:2018gcr,Benbrik:2019zdp,Cacciapaglia:2019zmj,
Aguilar-Saavedra:2019ghg,Zhou:2020byj,Wang:2020ips,Corcella:2021mdl,Cacciapaglia:2021uqh,Cui:2022hjg,Banerjee:2022izw,Banerjee:2022xmu,Bhardwaj:2022wfz,Bhardwaj:2022nko,Bardhan:2022sif}.
These new channels reduce the branching ratios into Standard Model final states, significantly alleviating current mass bounds~\cite{Banerjee:2024zvg}.  Compared with the LHC, clean collision environment and great development of detector techniques make it possible to detecting VLQs or provide complementary constraints on them at future lepton colliders. The Compact Linear Collider (CLIC) is a proposed TeV-scale  high-luminosity linear electron-positron collider which has a great potential for new physics detecting~\cite{Linssen:2012hp,CLIC:2018fvx}. Recently, searching for VLQs at the CLIC has drawn many attentions including pair production~\cite{CLICBB,Qin:2023zoi} and single production studies~\cite{Franceschini:2019zsg,Qin:2021cxl,Han:2021kcr,Han:2021lpg,Qin:2022mru,Han:2022exz,Han:2022zgw,Han:2022rxy,Yang:2023wnv}.

In this paper, we study the search of TeV-scale VLQ-$B$ quark via the pair production followed by the decays of $B\rightarrow bZ$ and $B\rightarrow bh$ at future Compact Linear Collider (CLIC)  with center-of-mass of 3 TeV.  We take the singlet VLQ-$B$ as a showcase and  focus on the subsequent fully hadronic decays of $Z$ and $h$ bosons which has a large branching ratio with the compensation for the relative small production rate. The large mass of VLQ-$B$ will induce the highly boosted bosons $Z$ and $h$, which tend to manifest as fat-jets. Clean collision environment and relative rare background events with large mass fat-jets at future lepton colliders inspire us to study this channel.

The structure of this paper is as follows. After summarizing the interactions  and decay modes of the singlet vector-like B quark in section II, we perform a detailed analysis on the probability of detecting vector-like $B$ via the process $e^+e^- \rightarrow B\bar{B}$ followed by  $B(\bar{B})\rightarrow bZ$ or $bh$ with subsequent hadron decay of $Z$ and $h$ at CLIC .  Finally, summaries and discussions are given in section IV.

\section{Effective interactions of VLQ-$B$}
The VLQ-$B$ are introduced in many new physics scenarios. Considering only the couples of VLQ-$B$ to the third generation SM quarks,  the effective Lagrangian for the singlet VLQ-$B$ is given by~\cite{Buchkremer:2013bha}:
\begin{equation}
\begin{split}
\mathcal{L}_B  =& \frac{g\kappa_B}{\sqrt{2}}\left\{\frac{1}{\sqrt{2}}\left[\bar{B}W_\mu^{-}
\gamma^{\mu}t_L\right]+\frac{1}{2\cos\theta_W}\left[\bar{B}Z_\mu^{-}\gamma^{\mu}
b_L\right]\right.\\
& \left. -\frac{m_B}{2m_W}\left[\bar{B}_RHb_L\right]-\frac{m_b}{2m_W}\left[\bar{B}_L
Hb_R\right]\right\}\\
&-\frac{e}{6\cos\theta_W}\{\bar{B}B_{\mu}\gamma^{\mu}B\}-
\frac{e}{4\sin\theta_W}\{\bar{B}W^{3}_{\mu}\gamma^{\mu}B\}+ h.c.,
\end{split}
\end{equation}
where $g$ is the $SU(2)_L$ gauge coupling constant and $\theta_W$ is the Weinberg angle.  The VLQ-$B$ mass $m_B$ and  the relative coupling strength $\kappa_B$ are regarded as free parameters.

For a heavy weak-isospin singlet VLQ-$B$, they are generally assumed to decay into a third-generation quark and either a $W/Z$ boson or a Higgs boson  and  the relationship of the branching ratios (BRs) of three standard decay modes is
\bea
Br(B \rightarrow bh) \approx Br(B \rightarrow bZ) \approx \frac{1}{2} Br(B \rightarrow tW).
\eea
This relationship is a good approximation as expected from the Goldstone boson equivalence theorem~\cite{He:1992nga,He:1993yd,He:1994br,He:1996rb,He:1996cm}.

The VLQ searches at the LHC genarally assume only three standard decay modes with relevant BR relationship or foucus a concrete decay channel with a 100\% BR. However, the VLQ can natrally decay into light particles in new physics senarios, such as $B\rightarrow bS$ where $S$ is an additional scalar or pseudoscalar particle~\cite{Bhardwaj:2022nko,Bardhan:2022sif}.  With the introduction of  new exotic decay modes $B\rightarrow X$, the relationships of BRs can be described as:
\bea
Br(B \rightarrow bh) +Br(B \rightarrow bZ) +Br(B \rightarrow tW)=1-\beta_{new}\\
Br(B \rightarrow bh) \approx Br(B \rightarrow bZ) \approx \frac{1}{2} Br(B \rightarrow tW)\approx (1-\beta_{new})/4
\eea
where $\beta_{new}$ is the BR for the new exotic decay channels.
In this case, the BRs of the standard modes can become smaller and the LHC search bounds on VLQ-$B$ can be significantly relaxed.  In next section, we foucus on the decay modes of $B\rightarrow bZ(bh)$ with the assumption of the existence of exotic decay mode.

\section{COLLIDER SIMULATION AND ANALYSIS}

\begin{figure}[htb!]
\begin{center}
\includegraphics [scale=0.6] {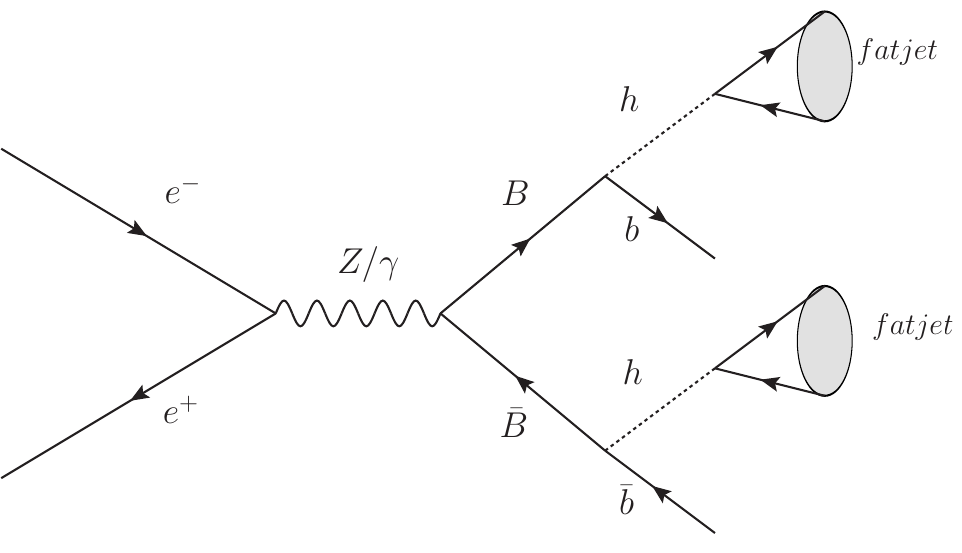}
\caption{Feynman diagrams for VLQ-$B$ pair production with subsequent decay into $bh$ at the CLIC. }
\label{fig:feynman}
\end{center}
\end{figure}

At 3 TeV CLIC, the VLQ-$B$ can be pair production with $\gamma$ or $Z$ mediation.  For a VLQ-$B$ with mass 1 (1.4) TeV,  the cross section $e^+ e^- \rightarrow B \bar{B}$ can reach 5.423 (3.075) fb at the CLIC with $\sqrt{s}=3 \text{TeV}$.  Unlike the QCD induced production at the LHC, the production rate of  VLQ-$B$ is not large. However, the rare background events, clean collision enviroment as well as the quick development of detector techniques make the CLIC be a hopeful machine to detecing the signature.
Considering both the decay modes of $B\rightarrow bh$ and $B\rightarrow bZ$ with subsequent fully hadronic decays of $Z$ and $h$ bosons, we analyze the prospect for detecing the VLQ-$B$. In these channels, the large branching ratio of hadronic decay can compenste for the relative small production rate. The large mass of VLQ-$B$ at TeV scale will induce  highly boosted bosons.  The hadronical decay products of a highly boosted bosons $Z$ or $h$ are more likely to form a fat-jet than two isolated jets. For these fat-jets, taking a large jet radius $R$ could capture all the decay products.  The Feynman diagram for the processes are demonstrated in Fig.~\ref{fig:feynman}.  It was found that searching for VLQ-$B$ in fully hadronic channels with a large R and jet substructure technique can overcome large combinatorics problem and efficiently suppress the large background events at the LHC~\cite{Yang:2014usa,Choudhury:2021nib}. We expect that the clean enviroment and rare boosted fat-jet background events make it possible to detect the VLQ-$B$ in the fully hadronic channels. In this section, we perform Monte Calo simulation and carry out phenomenological analyses in both boosted $h$ channel and boosted $Z$ channel, respectily. The parameter $\kappa_B=0.1$ and $Br(B\rightarrow bh)=Br(B\rightarrow bZ)=0.25$ is taken in the calculation.

In our  Monte Carlo study, the parton-level events  for the signals and SM backgrounds are generated by MadGraph5\_aMC\_v3.3.2~\cite{mg5} and then are interfaced to Pythia8~\cite{pythia8} for fragmentation and showering.  Subsequently, all event samples are fed into Delphes3~\cite{delphes} program to simulate detector effects. In our analysis, jets are clustered by employing FASTJET package~\cite{fastjet}. The Valencia Linear Collider (VLC) algorithm~\cite{Boronat:2014hva,Boronat:2016tgd} are taken with a large jet radius parameter $R$ = 1.0 to capture fat-jets. Finally, a cut-based analysis is performed by using MadAnalysis 5~\cite{ma5}.

\subsection{Boosted $h$ Channel}
In this subsection,we analyze the signal and background events at the 3-TeV CLIC via the process $e^+e^-\rightarrow  B\bar{B} \rightarrow bh\bar{b}h\rightarrow J_hJ_hb\bar{b} $ , where $J_h$ denote the boosted fat-jet from higgs decay.  It is expected that the highly boosted higgs jets come from the decay of TeV-scale VLQ-$B$ could be captured with a large jet radius $R=1.0$.

The dominant SM backgrounds mainly come from the processes:
\begin{itemize}
\item $e^+ e^- \rightarrow W^+ W^- Z$  with $Z/W^{\pm} \rightarrow j j$ .
\item $e^+ e^- \rightarrow t \bar{t}$ with $t(\bar{t}) \rightarrow W^{\pm} b$ and $W^{\pm} \rightarrow j j$.
\item $e^+ e^- \rightarrow t \bar{t} h$ with $t(\bar{t}) \rightarrow W ^{\pm}b$, $W^{\pm} \rightarrow j j$  and $h \rightarrow b \bar{b}$.
\end{itemize}
In following analyses, we don't take $b$-tagging which is not valid for the discrimination from backgrounds because many background processes have $b$-jets.
\begin{figure}[htb]
\begin{center}
\includegraphics [scale=0.27] {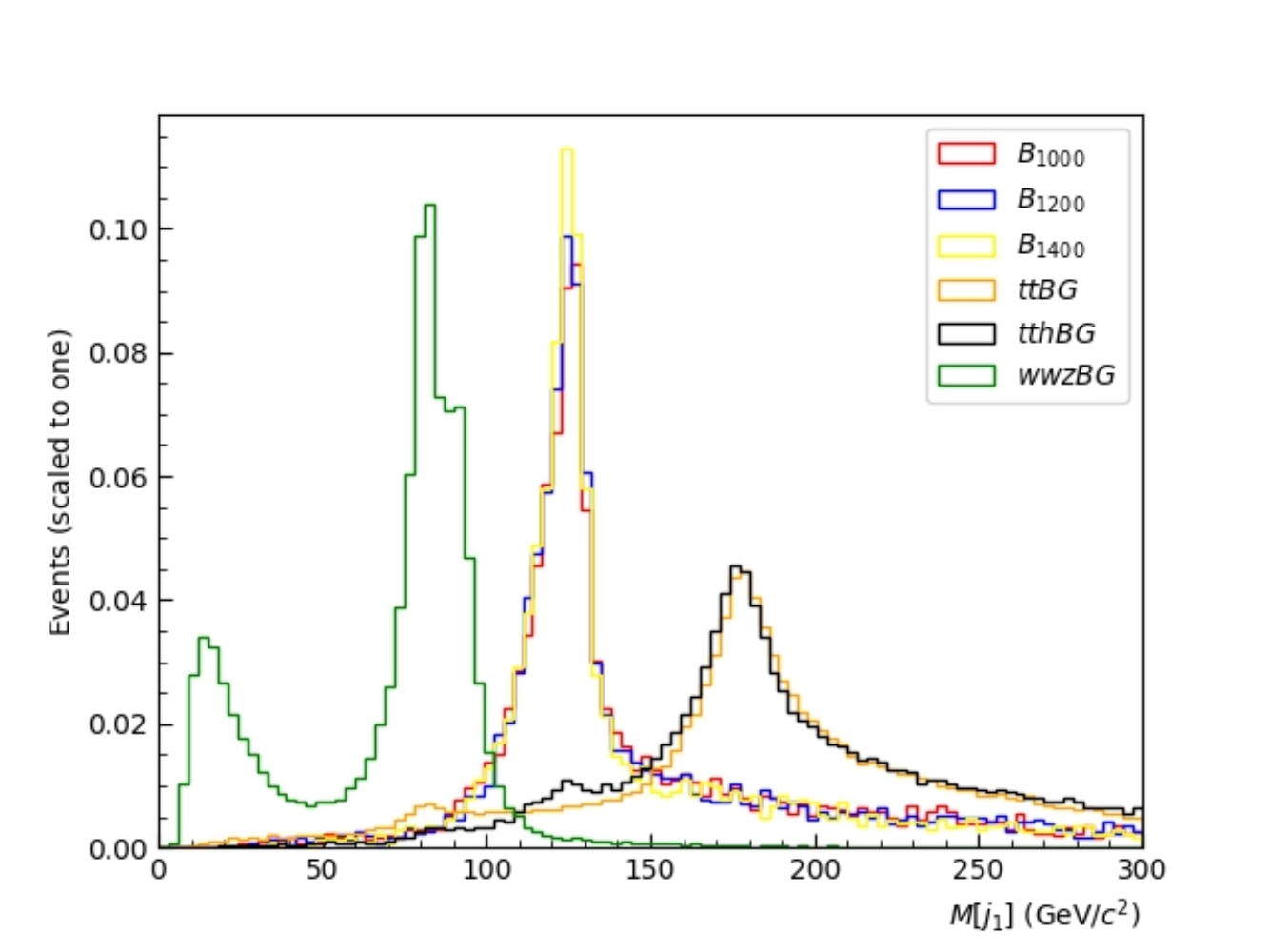}
\includegraphics [scale=0.27] {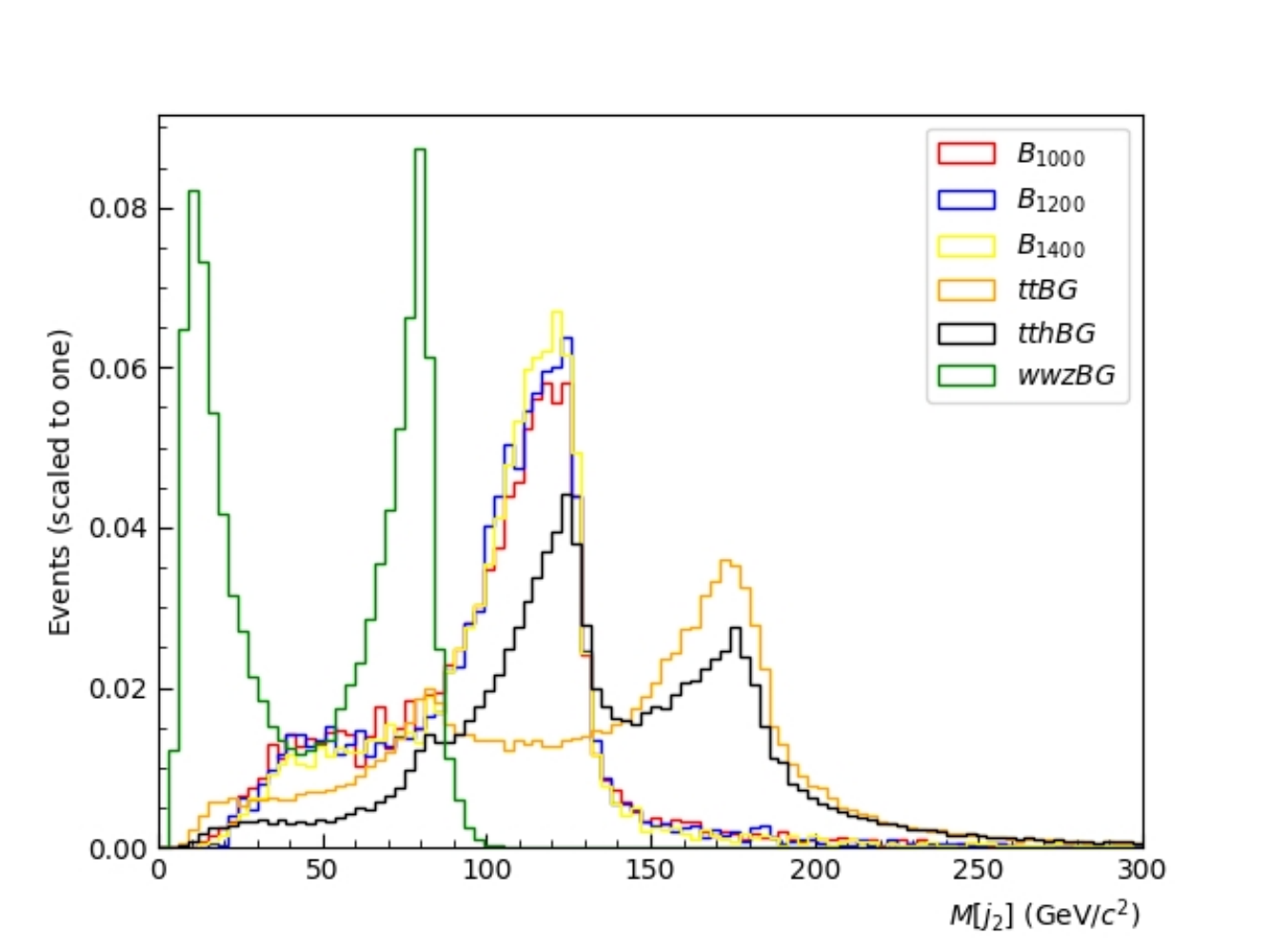}
\includegraphics [scale=0.27] {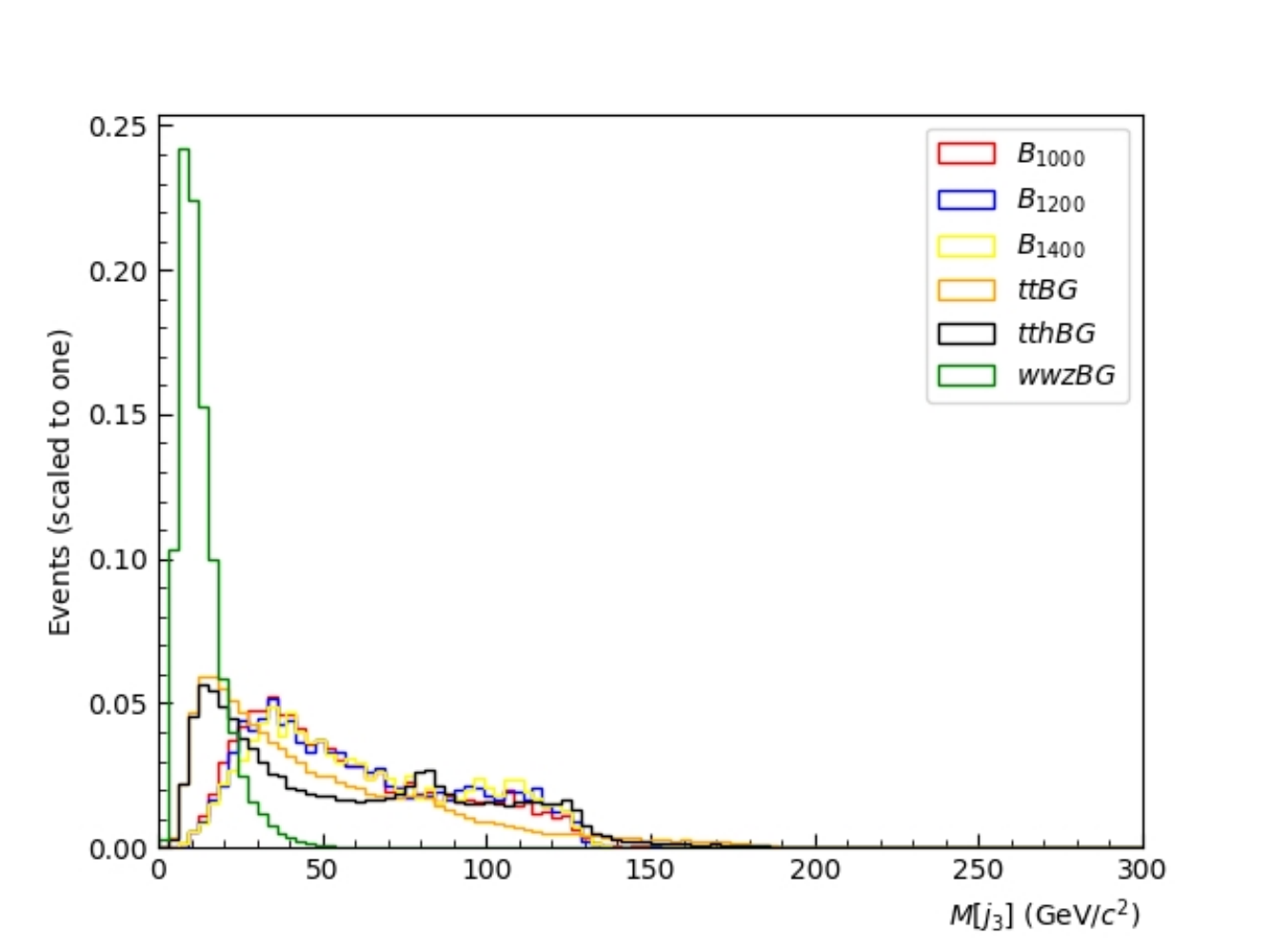}
\includegraphics [scale=0.27] {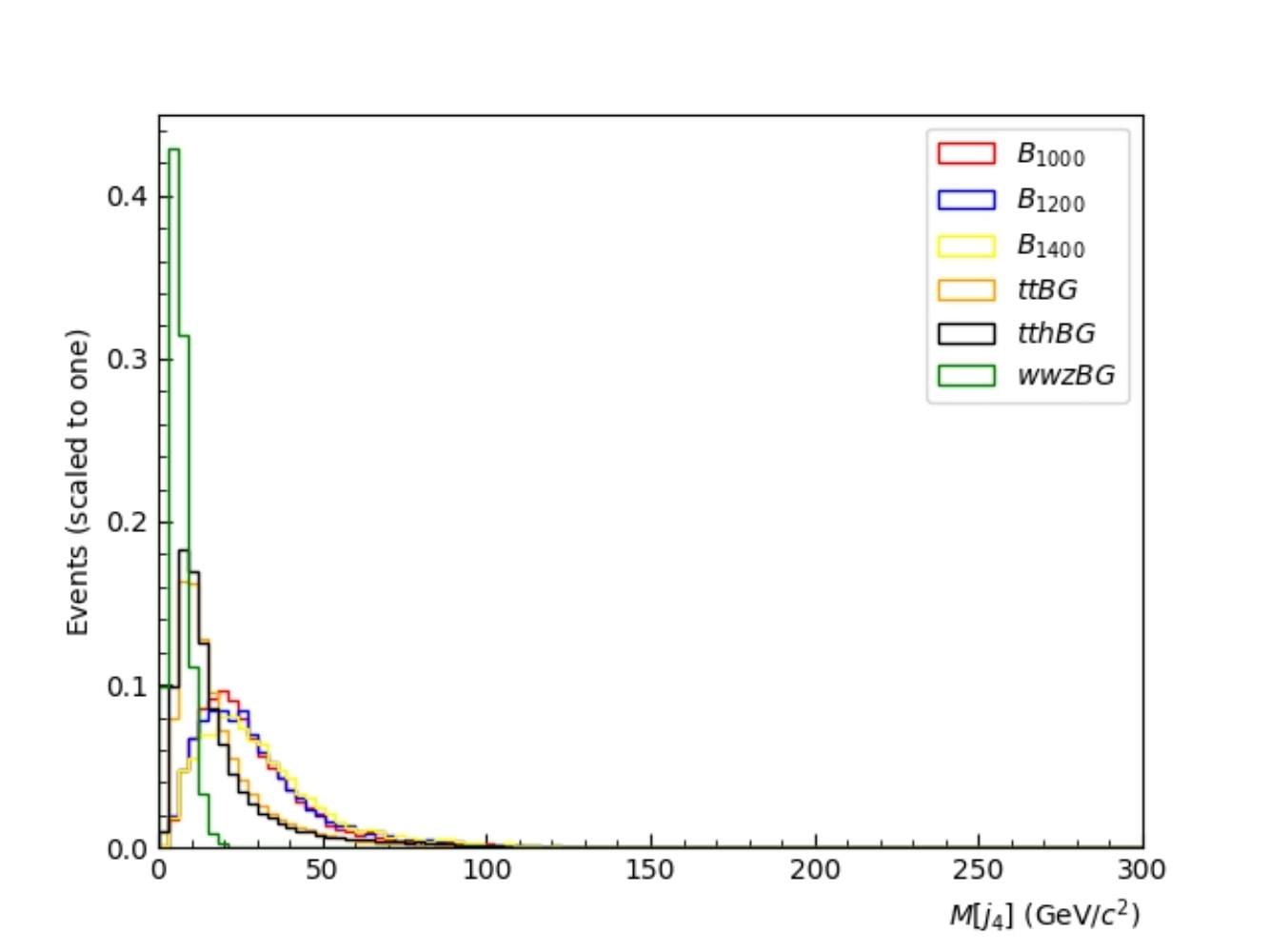}
\caption{The normalized distributions for jet mass of  for the signal and backgrounds.}
\label{fig:jetmassbh}
\end{center}
\end{figure}

\begin{figure}[!ht]
\begin{center}
\includegraphics [scale=0.27] {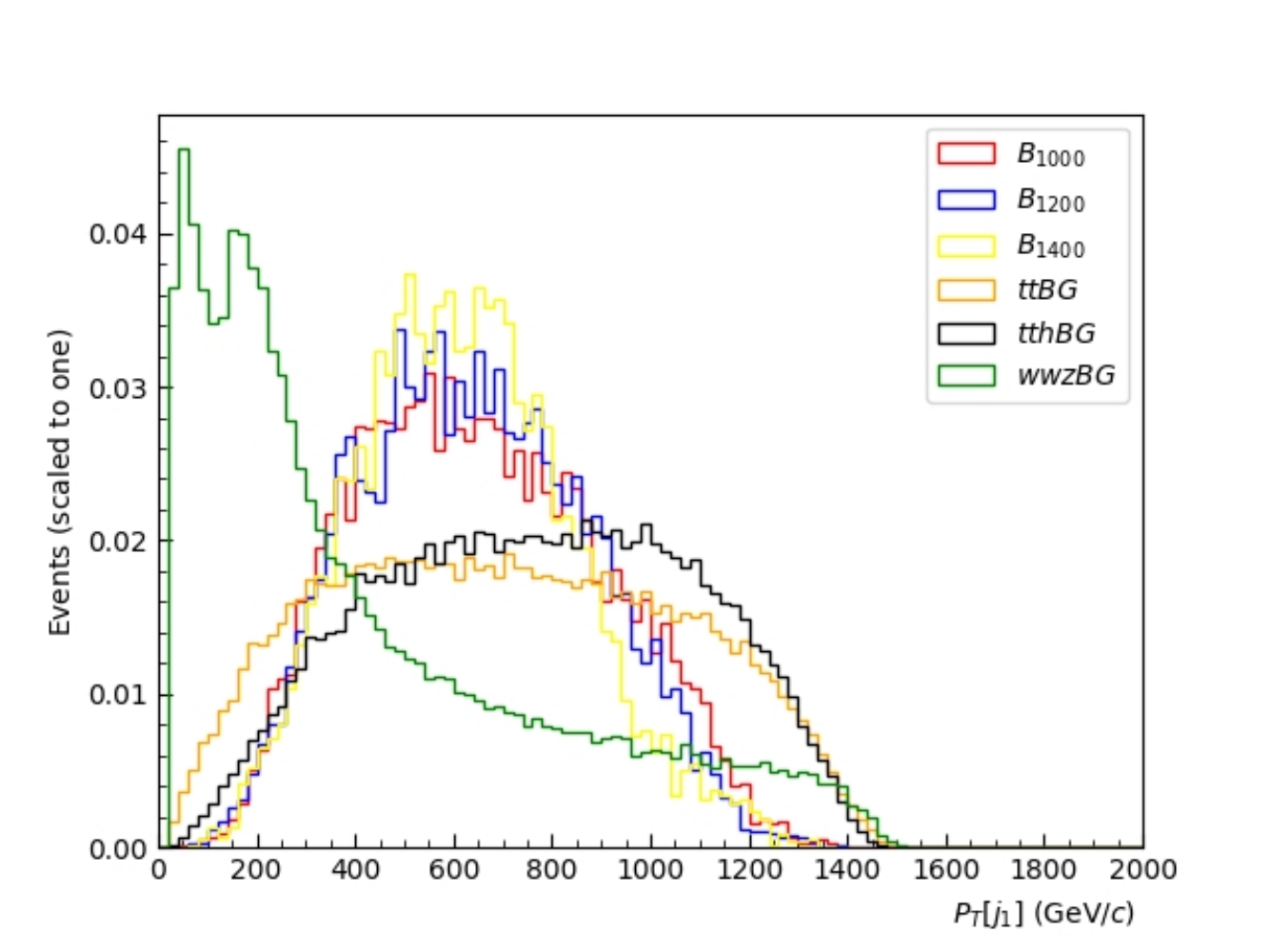}
\includegraphics [scale=0.27] {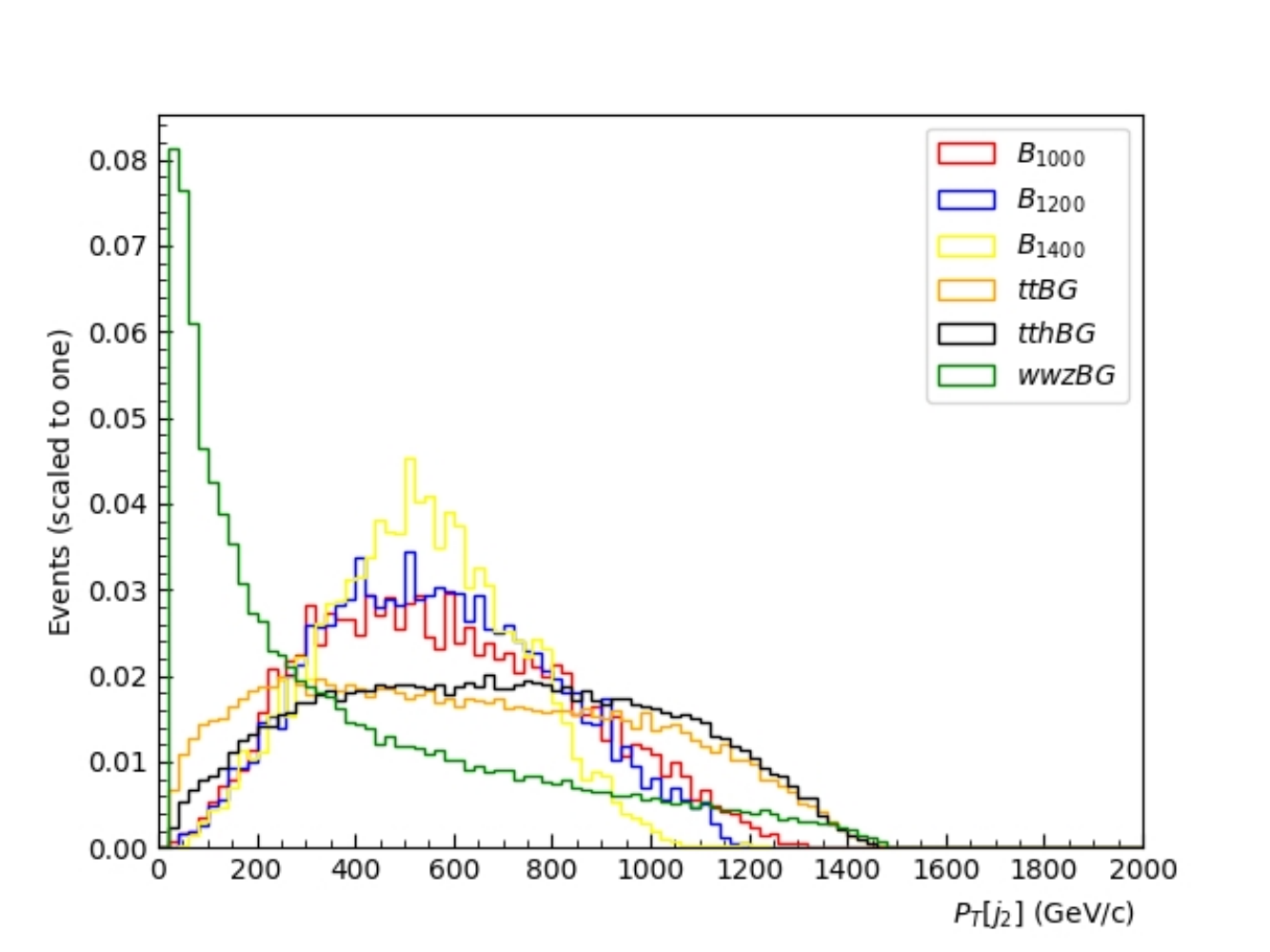}
\includegraphics [scale=0.27] {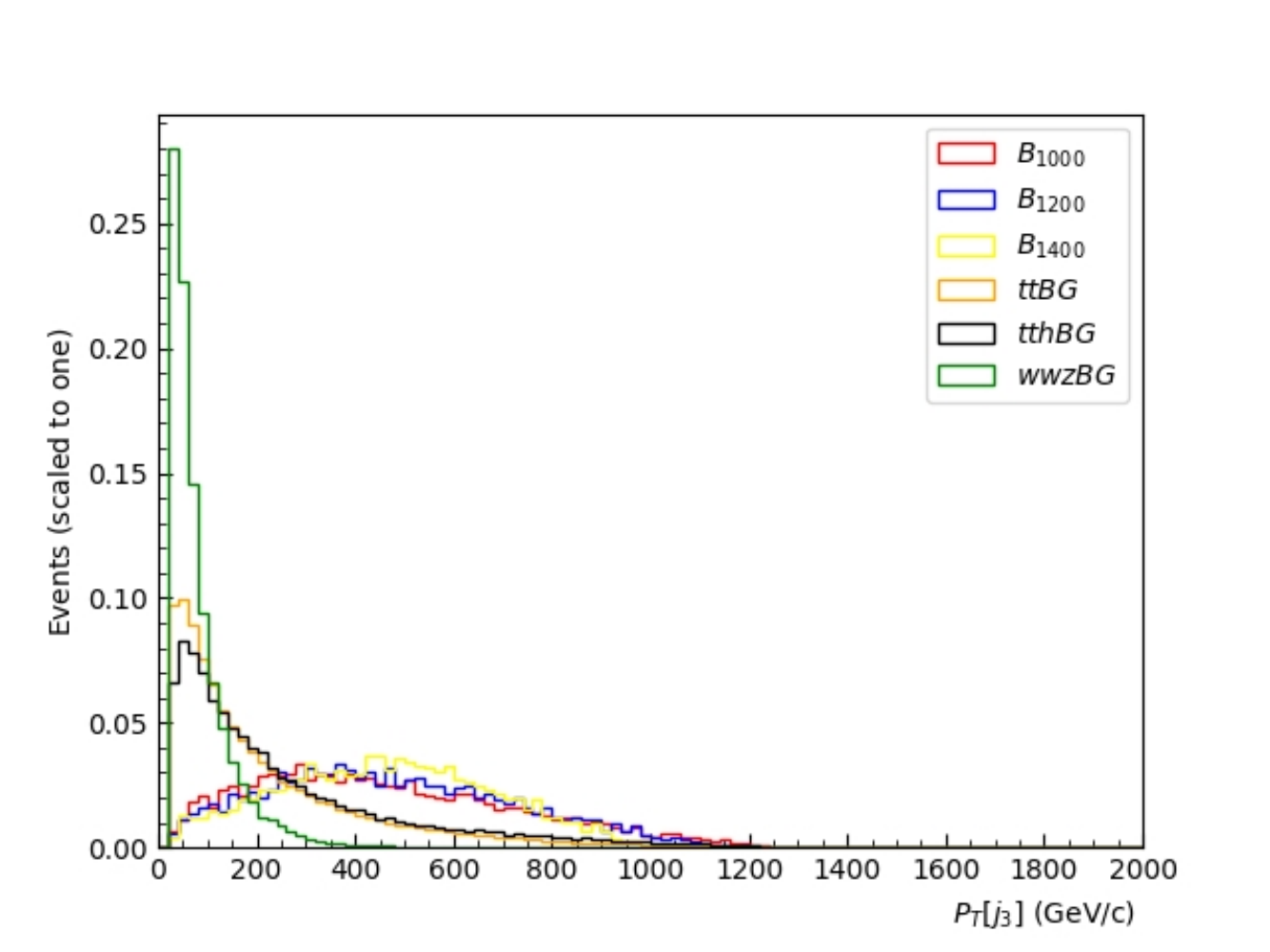}
\includegraphics [scale=0.27] {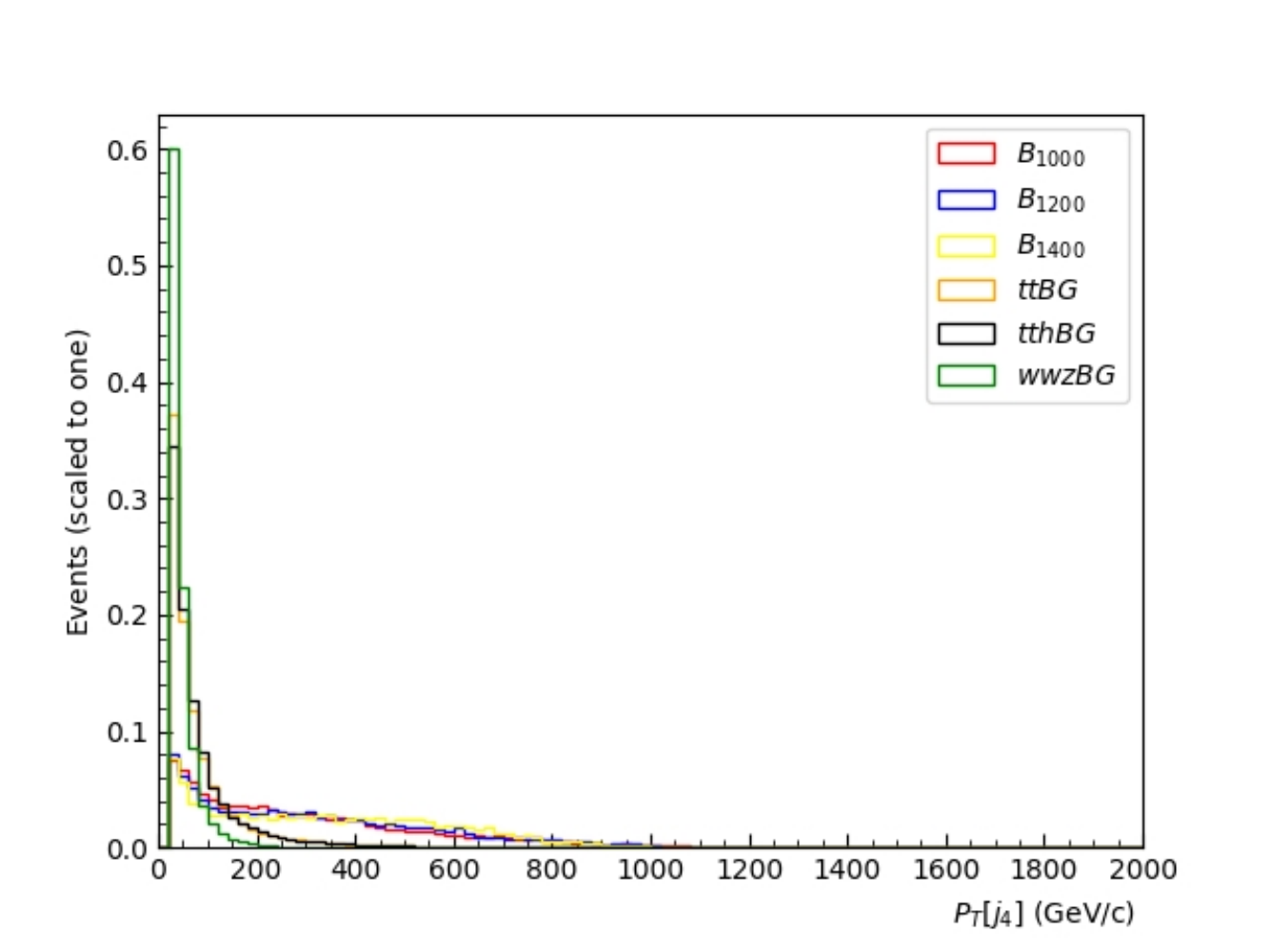}
\includegraphics [scale=0.27] {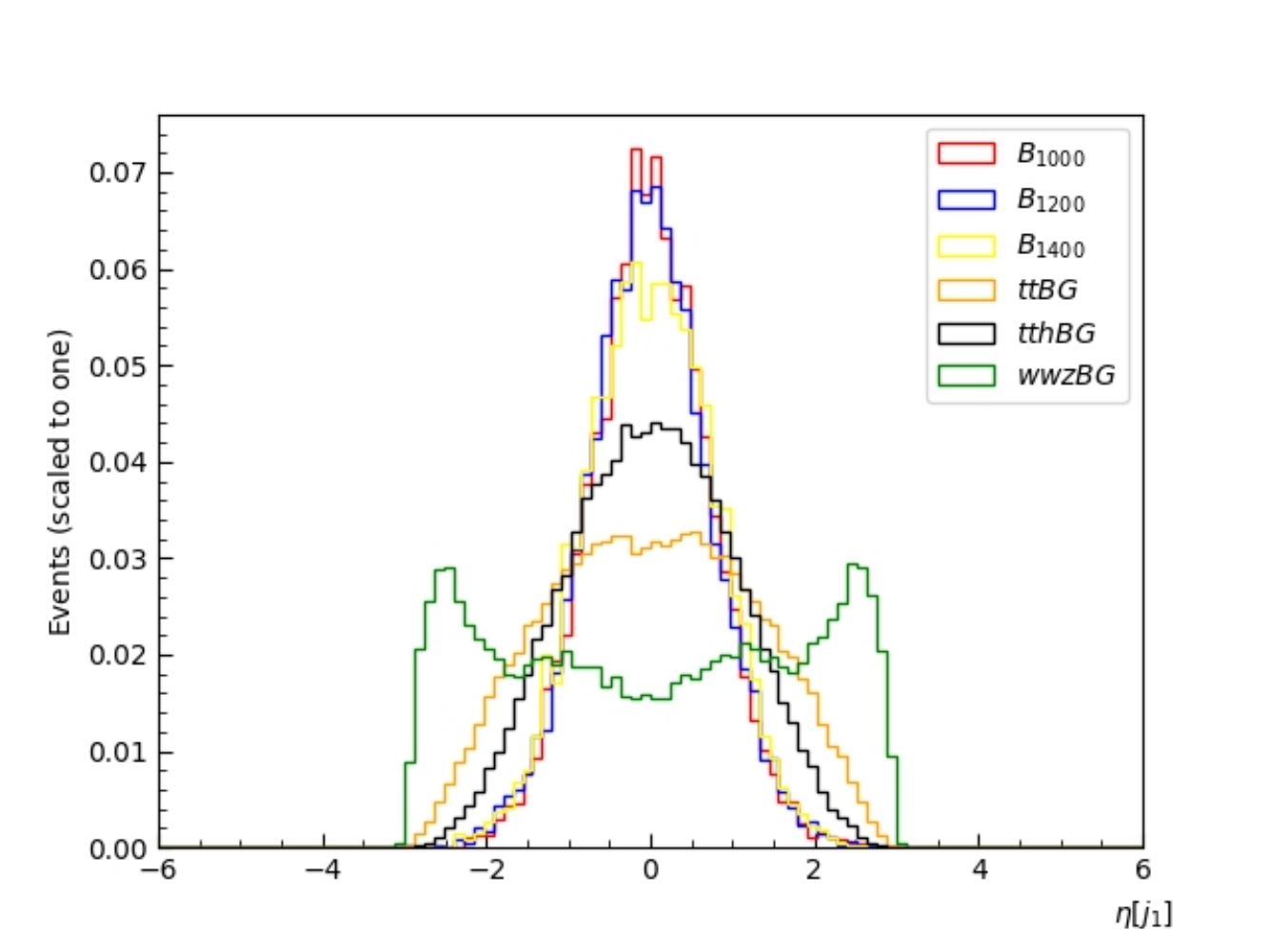}
\includegraphics [scale=0.27] {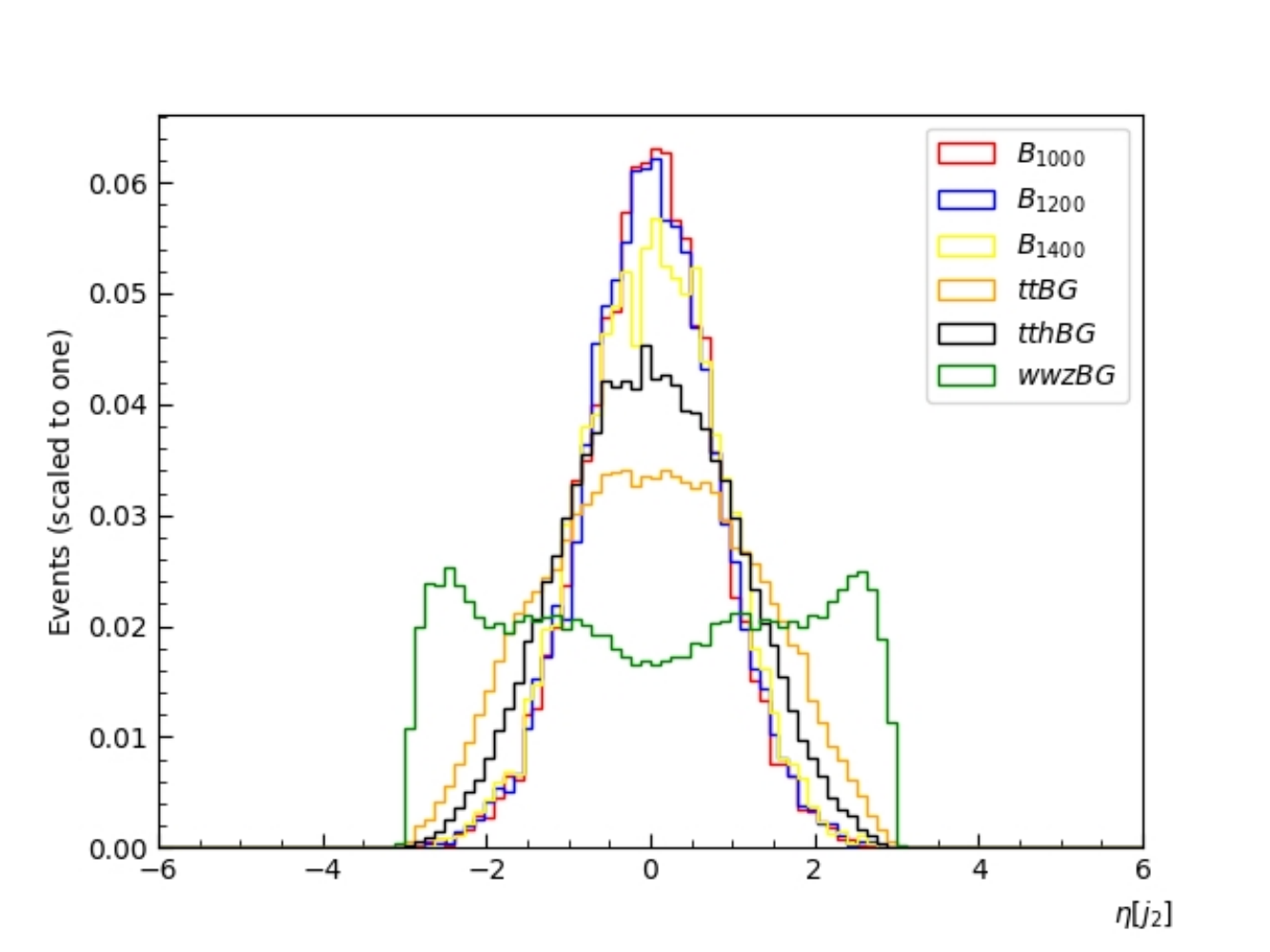}
\includegraphics [scale=0.27] {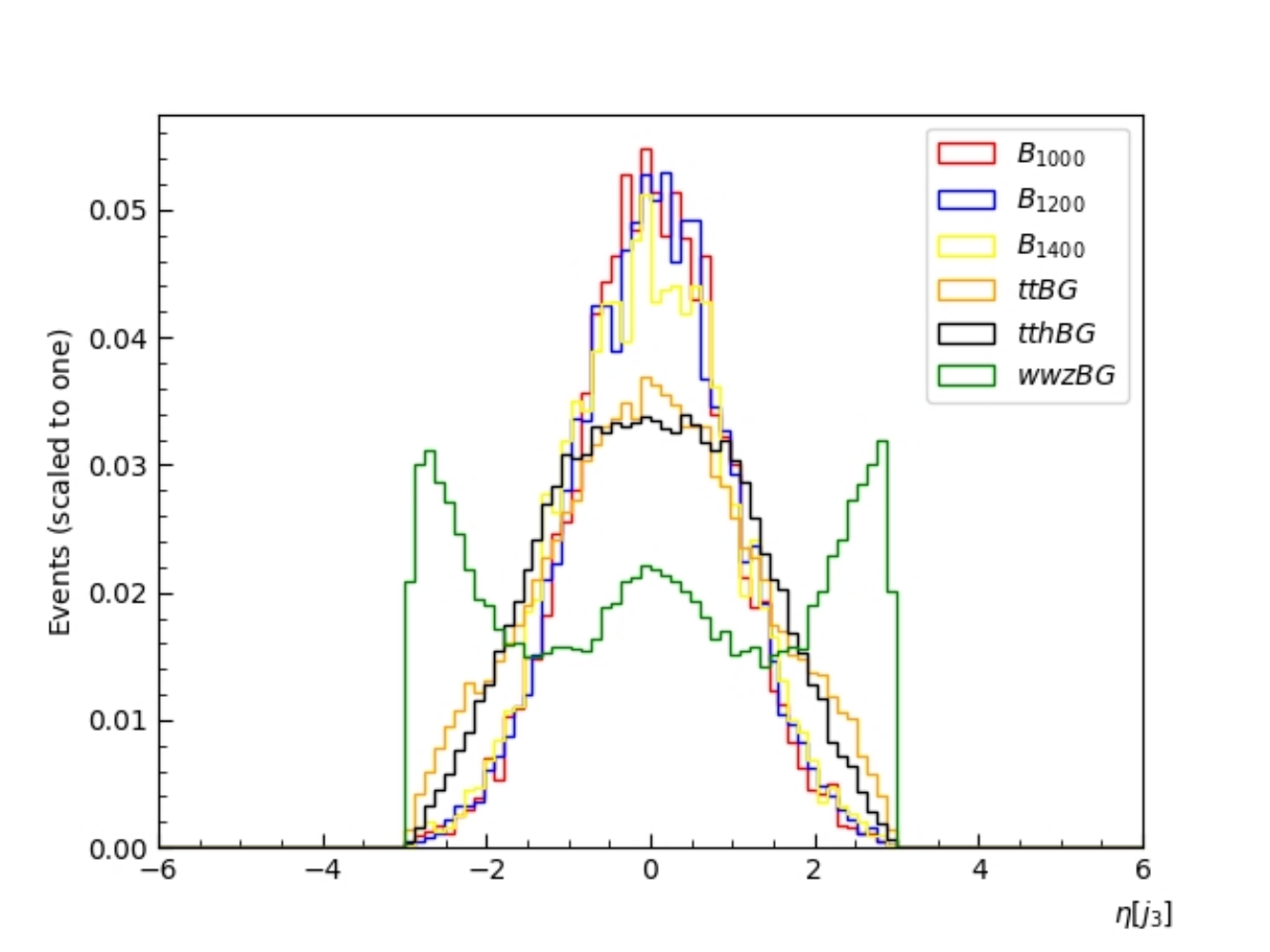}
\includegraphics [scale=0.27] {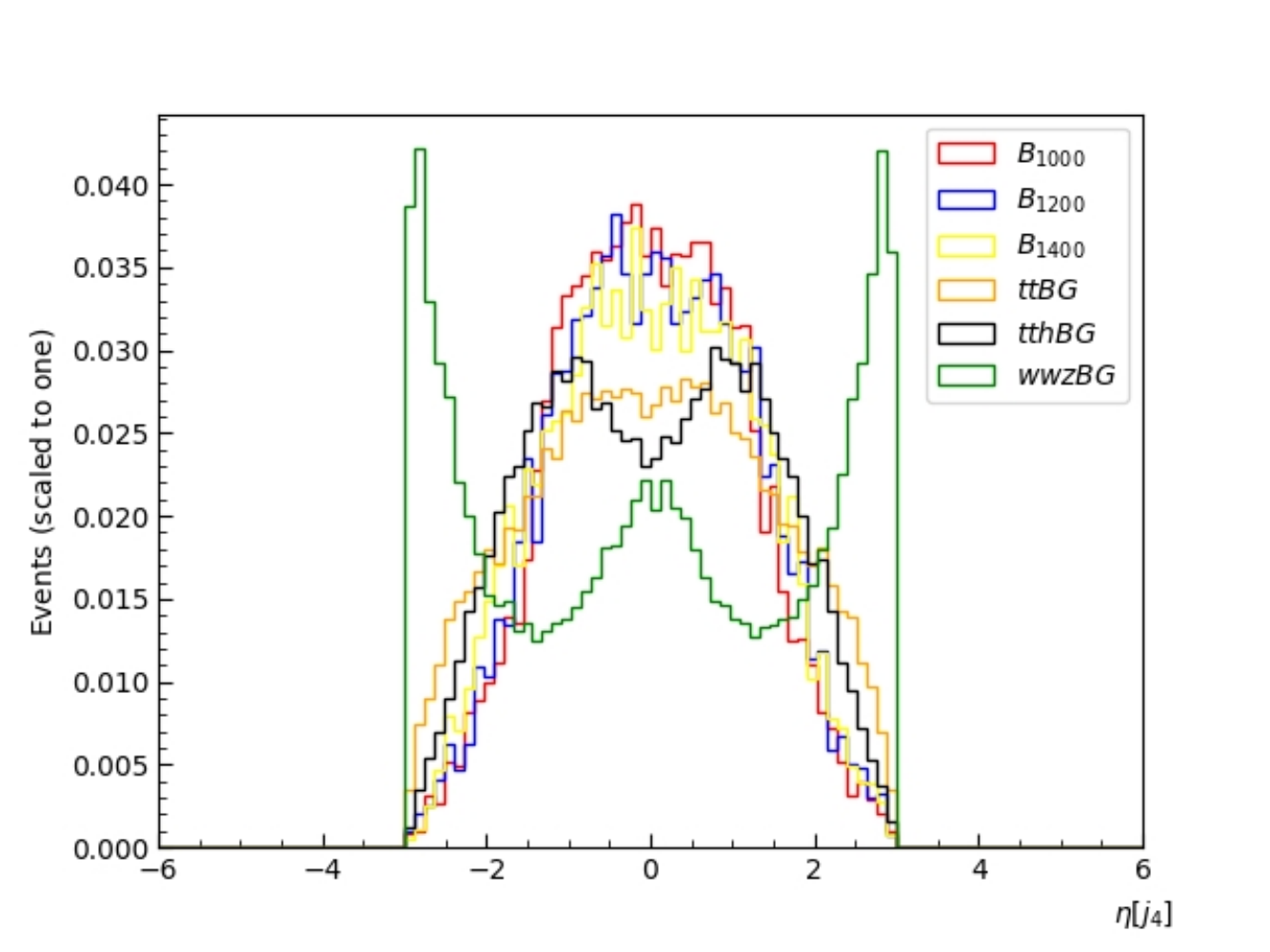}	
\caption{The normalized distributions for transverse momentum and pseudo-rapidity of jets for the signal in boosted higgs channel and backgrounds.}
\label{fig:jetptbh}
\end{center}
\end{figure}

To simulate detector acceptance and provide effective trigger, we choose the basic cuts at parton level for the signals and SM backgrounds as follows:
\begin{center}
$P_T(j)>20 \text{GeV},|\eta(j)|<3$
\end{center}
where $P_T(j)$ and $\eta(j)$ are the transverse momentum and pseudo-rapidity of jets. For signal events,  the heavy mass of  VLQ-$B$ transfer to the large transverse momentum of B's decay products  $b$ and $h$.  After selecting events with more than 4 jets in the final state, we resort the final jets with order of mass and present the normalized kinematic distributions for the signal and backgrounds in Fig. 2. For the signal, the hadronic decay products of the boosted higgs bosons generally can be captured by the large jet radius $R=1.0$. As shown in the figure, a large fraction of the signal events have two massive fat-jets with masses around the higgs mass $m_h$ and two isolated jets with a low mass. For the backgrounds, there also exist some events with boosted fat-jets due the large collision energy at the CLIC but the mass distributions of fat-jets are very different. Among the backgrounds,  it is difficult to find two massive jets around higgs mass.  The first massive jet of $t\bar{t}$ and $t\bar{t}h$ backgrounds are always a top-jet and only a tiny fraction of $t\bar{t} $ and $t\bar{t}h$ events could mimic two higgs-jets.  In addition, the transverse momentum of the massive jets of the signal peak around $m_B/2$ due to kinematic while the background has a wide distribution range.  Furthermore, the $WWZ$ background has very different distribution characters from the signal. Many $WWZ$ events have jets with large rapidity and small $P_T$ .

In order to extract the signal from the backgrounds, a set of improved cuts are adopted.
\begin{itemize}
\item Cut-1: There are at least four isolated jets and the scalar sum of all transverse momentum of all jets are required larger than 600 GeV, i.e,  $(N(j) \geq 4)$  and $H_T >$600.
\item Cut-2: The first two massive jets have mass in the higgs-mass window, i.e., $100<M_{j_{1,2}}<150$. Besides there are at least two light jets with mass requirement of $M_{j_{3,4}}<70$.
\item Cut-3: Each massive jet and a light have have a combined reconstructed mass greater than 300.
\end{itemize}

\begin{table}[tbh]
\begin{center}%
\begin{tabular}
[c]{|c|c|c|c|c|c|c|}\hline
\multirow{2}{*}{Cuts}    & \multicolumn{3}{c|}{Signals}  & \multicolumn{3}{c|}{Backgrounds}   \\
\cline{2-7}
& 1000GeV & 1200GeV & 1400GeV & $WWZ$ & $t\bar{t}$ & $t\bar{t}h$   \\ \hline
Basic  & 0.1314 & 0.118 & 0.07206 &  7.532 & 7.792 &0.1372\\
Cut-1  & 0.09796 & 0.0754  & 0.0397   &  0.5632  &  1.076   & 0.0401 \\ 
Cut-2  &  0.02197 & 0.0169  & 0.0088  &  0.000422  &  0.036 & 0.00208 \\
Cut-3  &  0.02005 & 0.0155  & 0.0082  &  0.000166  &  0.01196 & 0.00098 \\
$S/\sqrt{S+B}$&  7.8148  &   6.5169    & 3.9894    &      &     &                 \\
\hline	
\end{tabular}
\end{center}
\caption{Cut flow of the cross sections (in fb) for the signals with three benchmark masses of VLQ-$B$ quark and backgrounds in boosted higgs channel.  The statistical significance(SS) is calculated for an integrated luminosity of $5 \text{ab}^{-1}$}%
\label{table:bhbh}%
\end{table}

The cross sections for signals ($m_B = 1000, 1200,1400$ \text{GeV}) and  backgrounds are summarized in TABLE~\ref{table:bhbh}. One can see that all the SM backgrounds are suppressed very efficiently after imposing the cuts, while the signal events are kept in a relative large efficiency.  The statistical significance $S/\sqrt{S+B}$  is also presented in the table. When $m_B=1200$GeV,  with a luminosity of $5\text{ab}^{-1}$,  a significance of  6.5169  can be achieved.  For lighter VLQ-$B$ mass or a larger integrated luminosity, a good significance can be obtained.

\subsection{Boosted $Z$ Channel}

In this subsection, we consider the second possibility, namely when the heavy VLQ-$B$ decays in the mode of $B \rightarrow Z b$ with following decay $Z \rightarrow j j$. The topology of Feynaman diagram in this channel is the same as that in above subsection. In this case, the final state is $J_ZJ_Zb\bar{b}$ where $J_Z$ represent the fat jet derived from highly boosted $Z$ boson. For this signal, the dominant SM backgrounds come from the same processes as the case of boosted $h$ channel, i.e., $e^+ e^- \rightarrow W^+ W^- Z \rightarrow 6j$, $e^+ e^- \rightarrow t \bar{t} \rightarrow j jb j j\bar{b}$, and $e^+ e^- \rightarrow t \bar{t} h\rightarrow j j b j j \bar{b}b \bar{b}$.

\begin{figure}[htb]
\begin{center}
\includegraphics [scale=0.27] {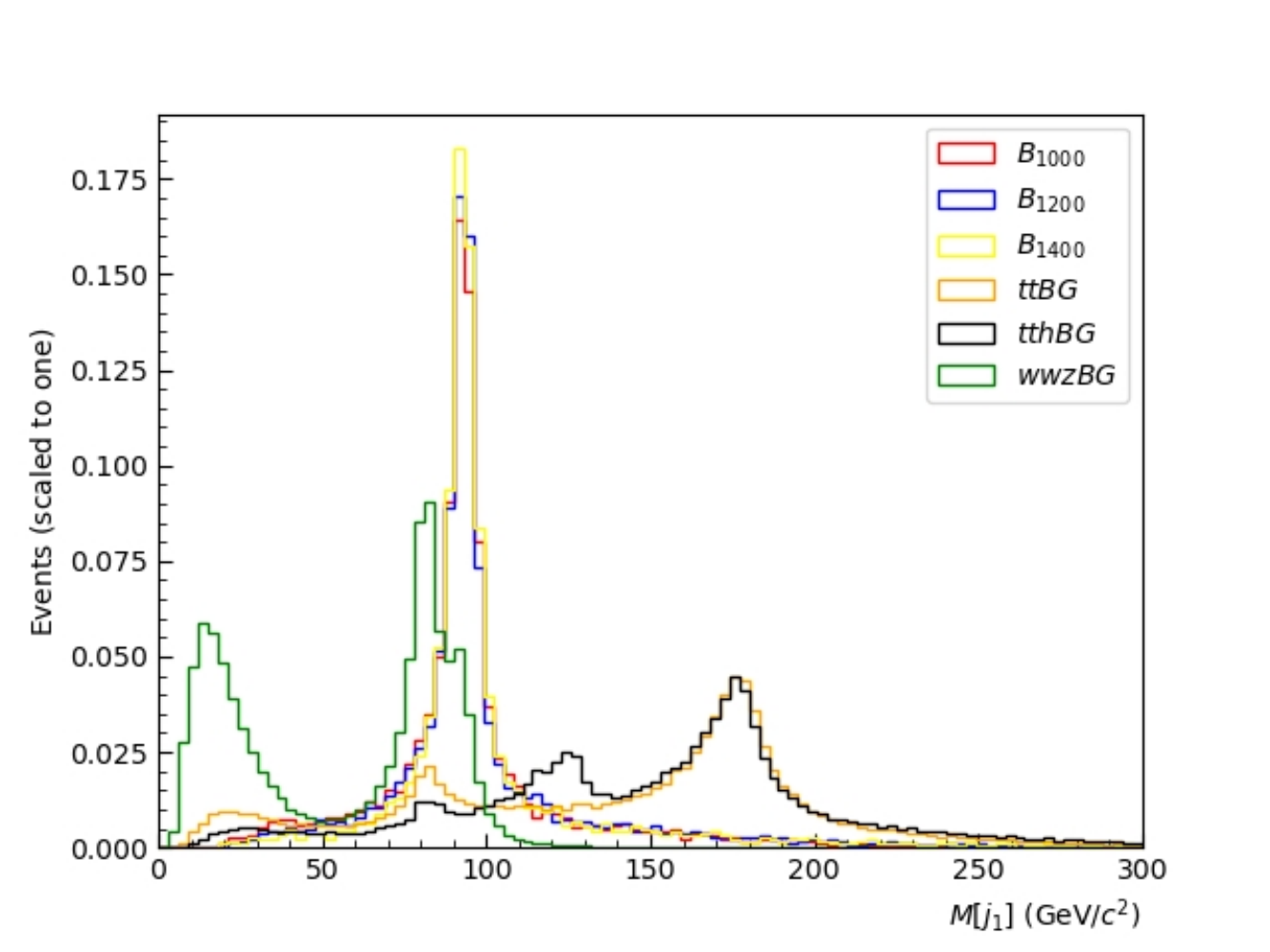}
\includegraphics [scale=0.27] {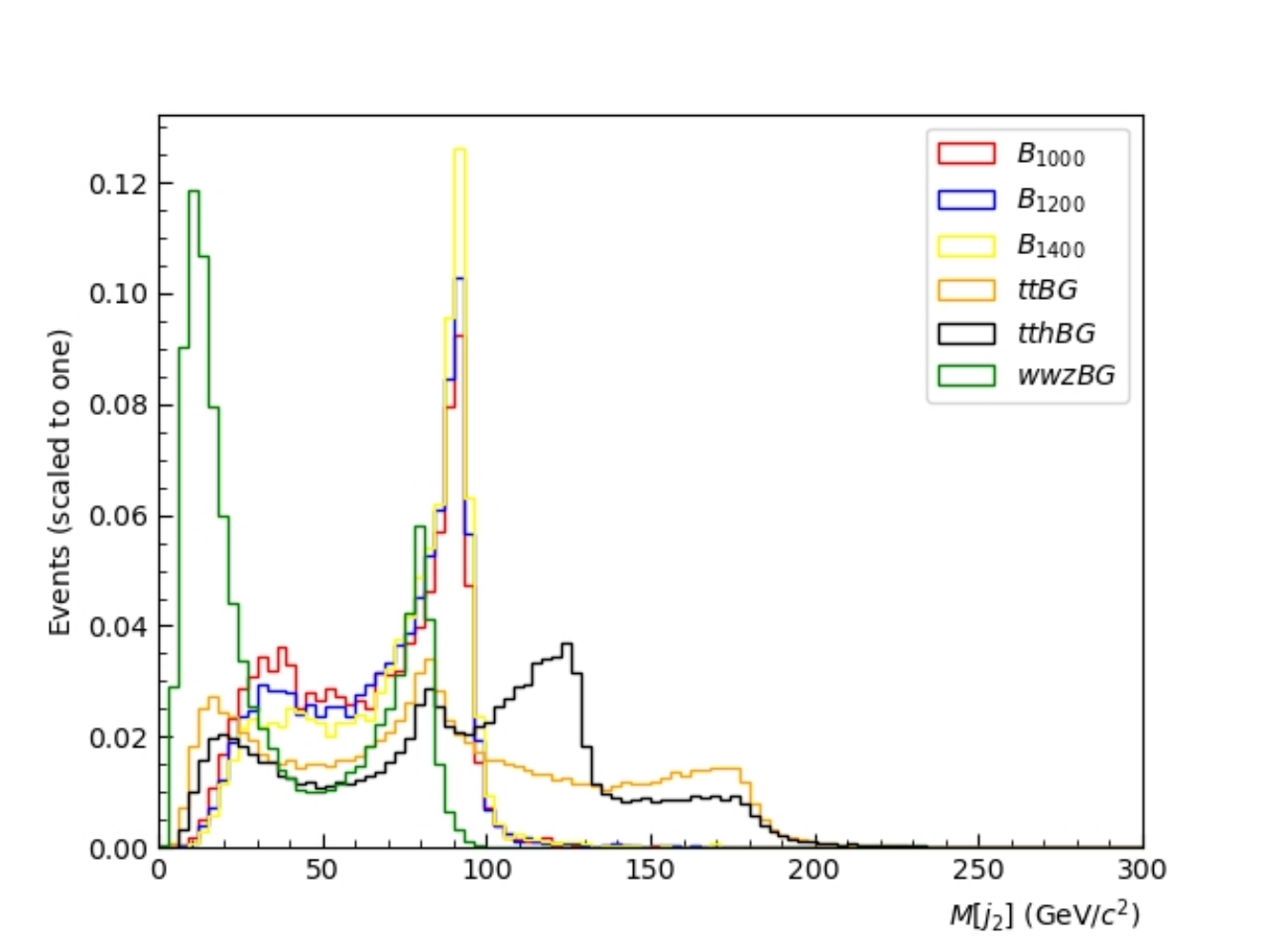}
\includegraphics [scale=0.27] {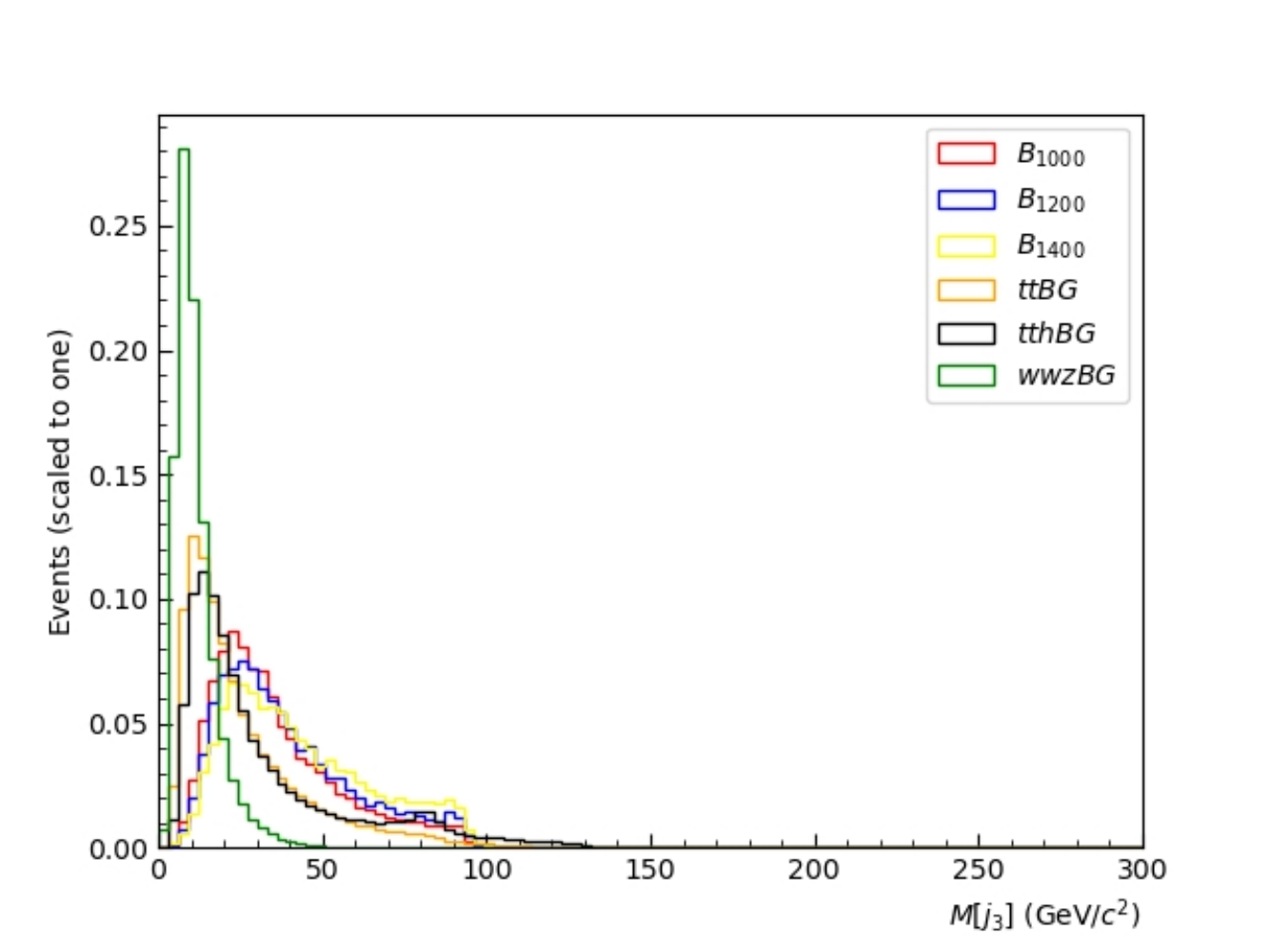}
\includegraphics [scale=0.27] {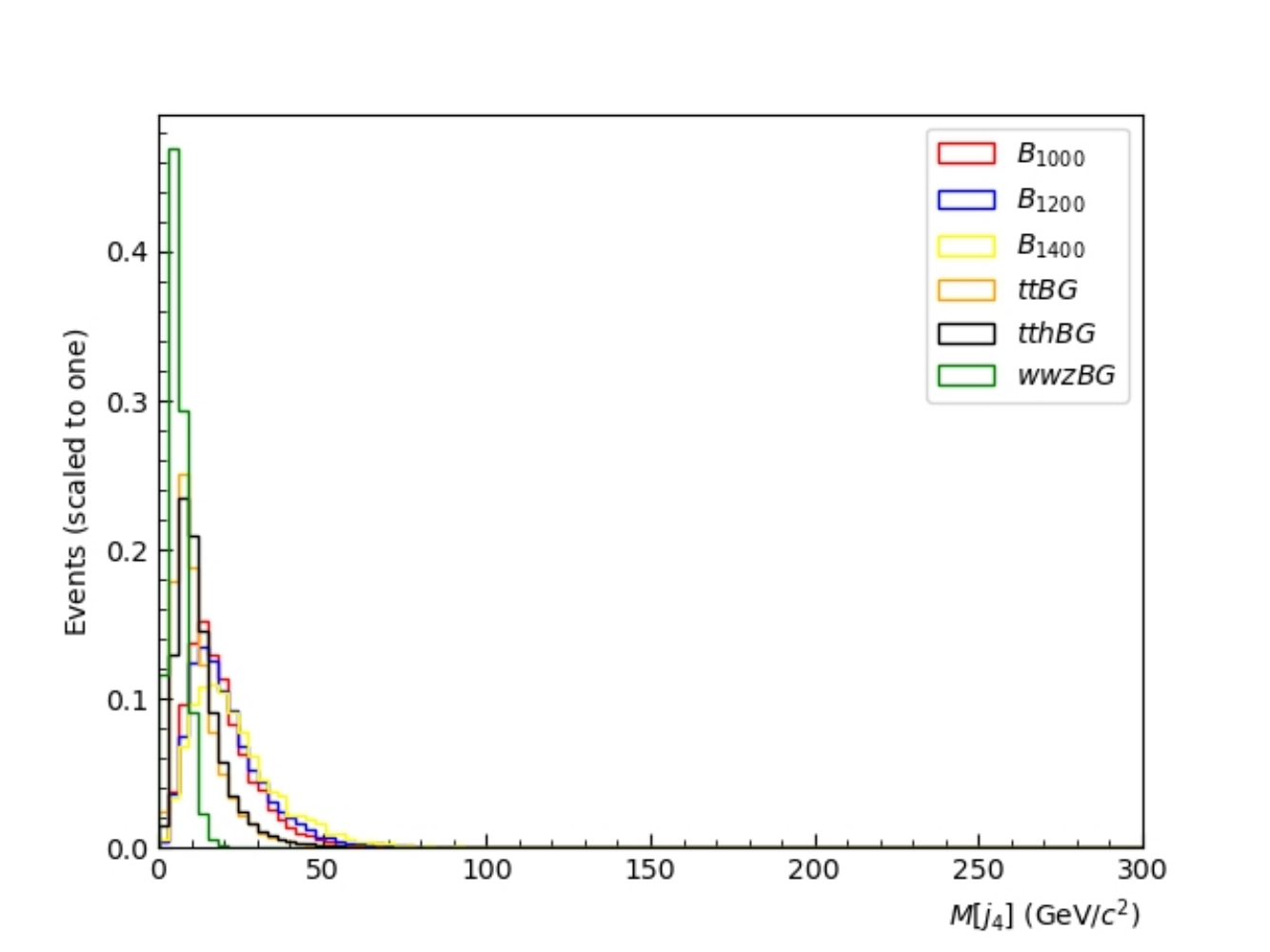}

\caption{The normalized distributions for the mass of first four  jets $M_{j_1}$, $M_{j_2}$, $M_{j_3}$ and $M_{j_4}$, respectively. }
\label{fig:jetmassbz}
\end{center}
\end{figure}

In this channel, we cluster the final hadrons with a large jet radius $R=0.8$ and choose the same basic cuts as those in above subsection.  Then we resort the jets with mass ordering.  The mass distributions for the mass ordering jets $j_{i}$ ($i=1\sim4$)  for the signal and backgrounds are presented in Fig.~\ref{fig:jetmassbz}. The distributions for transverse momentum and pseudo-rapidity in this boosted $Z$ channel are nearly the same to the case in boosted higgs channel, so we don't show them to avoid redundancy. The dominant background is $t\bar{t}$ due to its large production rate and decay topology. Although the mass peaks of the first two massive jets for the signal and the $WWZ$ background are overlap, most of  the $WWZ$  background could be cut-off  by a large $H_T$ cut. In order to  suppress the backgrounds, a set of  similar improved cuts are adopted as in above subsection:
\begin{itemize}
\item  Cut-1: There are at least four isolated jets and the scalar sum of all transverse momentum of all jets are required larger than 600 GeV, i.e,  $(N(j) \geq 4)$  and $H_T >$600.
\item  Cut-2: The first two jets have mass in the $Z$-mass window, i.e., $80<M_{j_{1,2}}<100$. Besides there are at least two light jets with mass requirement of $M_{j_{3,4}}<70$ .
\item  Cut-3: Each massive jet and a light have a combined reconstructed mass greater than 300.
\item  Cut-4: The combined reconstructed mass of the two light jets is greater than 100.
\end{itemize}

We present the cross sections for the signals with three benchmark masses ($m_B=1000, ~1200,~1400$ GeV) and the corresponding backgrounds in TABLE~\ref{table:bzbz}.  As shown in the table,  after a set of optimized cut, the backgrounds are suppressed efficiently. Taking $m_B=1200$GeV, a significance of 6.3358  for  a luminosity of $5\text{ab}^{-1}$ can be achieved. For heavier VLQ-$B$, the significance is smaller.

\begin{table}[tbh]
\begin{center}%
\begin{tabular}
[c]{|c|c|c|c|c|c|c|}\hline
\multirow{2}{*}{Cuts}    & \multicolumn{3}{c|}{Signals}  & \multicolumn{3}{c|}{Backgrounds}   \\
\cline{2-7}
& 1000GeV & 1200GeV & 1400GeV & $WWZ$ & $t\bar{t}$ & $t\bar{t}h$   \\ \hline
Basic  & 0.09796 & 0.08189 & 0.05245 &  7.532 & 7.792 &0.1372\\
Cut-1  & 0.09139 & 0.07474  & 0.04559   &  2.3128  &  3.355  & 0.1057 \\ 
Cut-2  & 0.01962 & 0.01753  & 0.01161  &  0.2564  &  0.0742 &  0.00209\\
Cut-3  & 0.0172 & 0.01533  & 0.01033  &  0.1005  &  0.0111 &  0.0005328\\
Cut-4  & 0.0169 & 0.01499  & 0.01018  &  0.002448  &  0.01001 &  0.000464\\
$S/\sqrt{S+B}$&  6.898  & 6.3358  &4.728  &      &      &                \\ \hline	
\end{tabular}
\end{center}
\caption{Cut flow of the cross sections (in fb) for the signals with three benchmark masses of VLQ-$B$ quark and backgrounds in boosted Z channel.  The statistical significance(SS) is calculated for an integrated luminosity of 5$ab^{-1}$}%
\label{table:bzbz}%
\end{table}

\begin{figure}[htb]
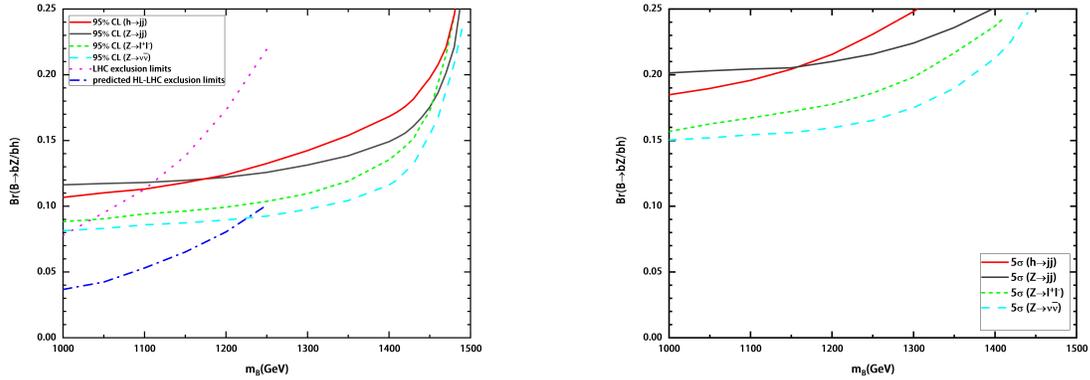

\begin{center}
\includegraphics [scale=0.035] {exclusionlimit}
\includegraphics [scale=0.035] {5sigma}
\caption{Exclusion limit (at $95\%$ CL) and discovery prospects (at 5$\sigma$) sensitivity reaches for boosted higgs channel and boosted $ Z$ channel at the 3 TeV CLIC with an integrated luminosity of $5\text{ab}^{-1}$ }
\label{fig:exclusion}
\end{center}
\end{figure}

\section{CONCLUSION}
With the introduction of exotic decay modes, the braching ratios of standard decay modes can be reduced and current limits from ATLAS and CMS can be relaxed. In this paper, considering two kinds of decay modes  $B\rightarrow bZ$ and $B\rightarrow bh$ ,  we have studied the search of TeV-scale weak-singlet VLQ-$B$ quark  in fully hadronic channel via the pair production at 3 TeV CLIC. The large mass of VLQ-$B$ will induce highly boosted bosons $Z$ and $h$, which tend to manifest as fat-jets.  We perform a full simulation for the signals in both boosted higgs channel and boosted $Z$ channel. After choosing a large jet radius and employing a set of cuts,  it is found that clean collision environment and rare fat-jets background events at the CLIC make it possible to extract the signals in some parameter space. Taking $m_B=1.2$ TeV ($1.4$ TeV) as an example, a signal significance  7.8148 (6.5169 ) can be achieved with an integrated luminosity of 5$\text{ab}^{-1}$ in booted higgs channel.  For the signal in boosted $Z$ channel,  a significance 6.898 (6.3358)  can be achieved with the same luminosity .

Here, we further consider the CLIC detecting sensitivity on the branching ratios Br($B\rightarrow bh$)  and $Br(B\rightarrow bZ)$.  In Fig.~\ref{fig:exclusion}, we plot the 95\% CL exclusion limit and 5$\sigma$ sensitivity reaches for boosted higgs channel and boosted $ Z$ channel at the 3 TeV CLIC with an integrated luminosity of $5\text{ab}^{-1}$ .  For the  boosted higgs  channel, the VLQ-$B$ can be excluded in the region of  Br($B\rightarrow bh$ )  $\in [ 0.107  , 0.25  ]$ and $m_B \in [ 1000  , 1480] $  GeV with an integrated luminosity of 5$ab^{-1}$.  For the  boosted $Z$ channel, the VLQ-$B$ can be excluded in the region of  Br($B\rightarrow bZ$ )  $\in [  0.116 ,0.25   ]$ and $m_B \in [ 1000  , 1490  ] $  GeV with an integrated luminosity of 5$ab^{-1}$.  As shown in the figure, fully hadronic channels and  leptonic channels~\cite{CLICBB}  at the CLIC  have similar performance in VLQ-$B$ searches. For comparison, we also present the observed 95\% CL exclusions limits at the 13 TeV LHC, the predicted exclusions at the future  HL-LHC with an integrated luminosity of $3000\text{fb}^{-1}$ , as well as the exclusions in leptonic channels from previous study~\cite{CLICBB}.  As shown in the figure, the future 3TeV CLIC with an integrated luminosity of $5\text{ab}^{-1}$ could provide better sensitivity than the current LHC searches  in the mass range of  $1120<m_B<1480$.  And even better sensitivity than the future HL-LHC in some large mass range. As for the discovery prospects, a VLQ-$B$ in the mass range from 1TeV to 1.4TeV  with Br($B\rightarrow bZ$>0.185) could be discovered at future 3TeV CLIC.

In addition, it is remarked that the analysis in this paper also applies to the moun collider with $ \sqrt{s}=3~\text{TeV}$.  For the 10 TeV muon collider, although the large phase space is opened, the cross section for pair production is significantly suppressed due to the s-channel suppression. For $m_B=1.5 \text{TeV} (2.5\text{TeV})  $, the cross section can only reach $  0.5345(0.5231)  \text{fb}$.  Generally speaking, it is not hopeful to detecting the TeV-scale VLQs at future 10 TeV muon collider due to limited production rates.

\section*{Acknowledgements}
This work was supported in part by the National Natural Science Foundation of China under Grants No. 12147214 and No. 11905093,  and the Basic Research Project of Liaoning Provincial Department of Education for Universities under Grants  No.~LJKMZ20221431 and Teaching Reform Research Project for graduates of Liaoning Normal University.

\end{document}